\documentstyle[12pt]{article}

\newcommand{\bb}{\begin{equation}}
\newcommand{\ee}{\end{equation}}
\newcommand{\bqn}{\begin{eqnarray}}
\newcommand{\eqn}{\end{eqnarray}}

\newcommand{\w}{\mbox{\tiny $\wedge$}}
\newcommand{\spc}{\mbox{\hspace{1cm}}}

\title{Darboux coordinates for (first order) tetrad gravity}

\author{M\'aximo Ba\~nados$^{1,2}$ and Mauricio Contreras$^{2,3}$}

\date{}

\begin{document}

\maketitle

\begin{center}

$^1${\em Departamento de F\'{\i}sica, Universidad de Santiago de Chile, Casilla 307, Santiago Chile.}\\
$^2${\em Centro de Estudios Cient\'{\i}ficos de Santiago,
Casilla 16443, Santiago, Chile. \\
$^3$Departamento de F\'{\i}sica, Facultad de Ciencias, Universidad de
Chile, Casilla 653, Santiago, Chile.} \\
\end{center}

\begin{abstract}
The Hamiltonian form of the Hilbert action in the first order tetrad formalism is examined.  We perform a non-linear field redefinition of the canonical variables isolating the part of the spin connection which is canonically conjugate to the tetrad. The geometrical meaning of the constraints written in these new variables is examined.  \\
\end{abstract}

\section{Introduction}

The tetrad formulation of general relativity is unavoidable if one wants to incorporate spinors into gravitational theories.  Among the many different ways to write the action in the tetrad formalism, the most attractive one is the so-called first order formalism in which the spin connection and tetrad fields are independent variables. In this formulation, the action looks like a gauge theory for the Poincare group. However, a closer analysis shows that only Lorentz rotations are symmetries of the action: the translational part of the Poincare group does not leave the action invariant \cite{GRG}. 

Despite the fact that gravity is not a gauge theory, there exists one modification of general relativity that resembles very much a Yang-Mills theory. This is the Ashtekar formulation which is based on a self-dual (complex) connection \cite{Ashtekar}. The Ashtekar formulation has stimulated a enormous effort towards the understanding of complex gravity in the Loop representation \cite{loop}, as well as to the application of Ashtekar ideas to the original, real, gravitational action (see \cite{Peldan} for a  review).   

In this paper we shall explore the first order formulation of tetrad gravity and its Hamiltonian structure.  The main goal of this note is to exhibit the existence of a non-linear field redefinition of the canonical variables that isolate the part of the spin connection which is canonically conjugate to the tetrad. This transformation can thus be interpreted as a Darboux transformation that brings the kinetical term in the action to a diagonal form, as discussed in \cite{Faddeev-Jackiw}.  The role of the constraints and their geometrical meaning are also discussed.

\section{Tetrad (first order) gravity}

We start with the Einstein-Hilbert action in four dimensions 
written in the tetrad (first order) formalism:
\bb
S[\omega,e] = \int_M \epsilon_{a b c d} R^{a b} \w e^{c} \w e^{d}, 
\label{action/0}
\ee 
where the curvature 2-form is defined by $R^{a b} = d \omega^{a b} +
\omega^{a }_{\ c} \w \omega^{c b}$ and $M$ is a four dimensional Hausdorf compact manifold. In writting (\ref{action/0}) the metric is not necessary because $R^{ab}$ is a 2-form and $e^a$ a 1-form. In this approach, the metric is a derived concept related to the tetrad by $g_{\mu\nu} = e^a_\mu e^b_\nu \eta_{ab}$. Here the indices $a,b,c,...$ run over the Lorentz group and $g_{\mu\nu}$ has thus Lorentzian signature.   

The action (\ref{action/0}) has two set of gauge symmetries. These are local Lorentz rotations and spacetime diffeomorphisms. One thus expect to find, in the Hamiltonian formalism, two sets of constraints which will generate these symmetries 

The first step towards a Hamiltonian formalism is to decompose spacetime
into space+time.  If the manifold has the topology $\Sigma
\times \Re$, the 1-forms $ \omega$ and $e$ can be decomposed in the
3+1 form,
\bqn
e^{a} &=& e^{a}_{0} dt + e^{a}_{i} dx^{i}, \\
\omega^{a b} &=& w^{a b}_{0} dt + \omega^{ab}_{\ \ i} dx^{i},
\eqn
where the $x^i$ are local coordinates on $\Sigma$ and $t$ runs along
$\Re$. 

Using this decomposition, up to surface terms \cite{RT}, the action
(\ref{action/0}) can be written in the form:
\bb
S=\int dt \int_{\Sigma}(\dot e^{a}_{i} \, \Omega^{\, i\  j}_{a\ bc} \,
 \omega^{bc}_{j}+\omega^{a b}_{0} J_{a b} + e^{a}_{0} P_{a}) 
\label{action/1}
\ee
where
\bqn
J_{a b} &=& \epsilon^{i j k}\epsilon_{a b c d}\, T^{c}_{ij} e^d_k, 
\label{Jab} \\
P_{a} &=& \epsilon^{ijk}\epsilon_{abcd}\, R^{bc}_{\ i j} e^d_k , 
\eqn
($T^{a}_{ij}$ and $R^{ab}_{ij}$ are the spatial components of
the torsion and curvature 2-forms respectively), and
$\Omega^{\, i\  j}_{a\ bc}$ is given by, 
\bb
\Omega_{a\ bc}^{\, i\  j} = 2 \epsilon_{abcd} \epsilon^{ijk} e^d_k.
\label{Omega} 
\ee

The variation of the action (\ref{action/1}) with respect to $e^a_0$ and $w^{ab}_0$ gives the equations,
\bb
P_a=0, \spc   J_{ab}=0,
\label{const0}
\ee
which do not contain any time derivatives. They are thus constraints
over the initial data.  As it can be expected, $P_a$ generates spacetime diffeomorphisms while $J_{ab}$ generates local Lorentz rotations.  However, to actually prove this assertion, we first need to study the symplectic structure encoded in the kinetical term of the action (\ref{action/1}).  

The combination $\Omega^{\, i\  j}_{a\ bc}$ appearing in (\ref{action/1})  is a $12\times 18$ matrix and therefore it cannot be inverted. [The number of components of the tetrad ($e^a_i$) is 4$\times$3=12, while the number of components of the spin connection ($w^{ab}_i$) is 6$\times$3=18].   Had $\Omega$ be non-degenerate, then the symplectic structure could have been easily derived following the formalism of  \cite{Faddeev-Jackiw}. In 2+1 dimensions, for example, this matrix is indeed non-degenerate and one finds that the triad and spin connection are conjugate pairs. In 3+1 dimensions (or higher), $\Omega$ fails to be invertible and one cannot infer the Poisson bracket structure in a straightforward manner. Note also that $\Omega$ is not only a rectangular matrix but it also depends explicitly on the tetrad which is a canonical coordinate.  

The number of zero eigenvalues of $\Omega$ is at least 6 [since it has 12
rows and 18 columns].  The corresponding six null vectors have the simple form,          
\bb 
U^{a b\ (kl)}_{\, k} = \delta^{(k}_k \epsilon^{l)mn} e^a_m e^b_n,    
\label{U}
\ee 
Indeed, it is easy to check the identity,
\bb 
\Omega^{\, i \ \, j}_{a\ bc} U^{bc\ (kl) }_{j} = 0. 
\label{OU} 
\ee 
The parenthesis $(kl)$ denotes symmetrization. Since the indices
$k,l$ take the values $1,2,3$, the combination $U^{a b\ (kl)}_{\, k}$ indeed represent six null vectors of $\Omega$.   

One can now prove that $\Omega$ does not have any further
zero eigenvalues by noticing that the $18\times 12$ matrix, 
\bb
\Theta^{a b \ \, c}_{\, i \ \ j} := 
\frac{-e^{[a}_{i} n^{b]} e^{c}_{j}
            + e^{[a}_{i} e^{b]}_{j} n^{c}
            +2 e^{[a}_{j} n^{b]} e^{c}_{i} }{8 \sqrt{g}}
\label{Theta}
\ee
where the square brackets indicate antisymetrisation, $g$ is the
determinant of the induced metric $g_{ij}=e^a_i e^b_j \eta_{ab}$, and
$n^{a}$ is the normal defined by,
\bb
n_{a} = \frac{1}{6 \sqrt{g}} \epsilon_{abcd} \epsilon^{ijk} e^{b}_{i}
e^{c}_{j} e^{d}_{k},
\label{normal}
\ee
($n^{a} e_{a i}= 0$ and $n^{a} n_{a} = -1$) is the right inverse of $\Omega$. Indeed, by direct calculation one can check the $\Theta$ and $\Omega$ satisfy, 
\bb
\Omega^{\, i\ \,j}_{a\ bc} \Theta^{bc\ d}_{\, j\ \,k} = \delta^i_k
\delta^d_a.
\label{OT}
\ee
 
The pair $\Theta$ and $U$ describe a basis of vectors in the 18 dimensional vector space $(^{ab}_i)$. This will be the key observation that will allow us to diagonalise the kinetical term of (\ref{action/1}). 

\subsection{Darboux coordinates}

As stressed above, the action (\ref{action/1}) is not yet in canonical form because the kinetical term does not give rise to an invertible 2-form on phase space. There are two equivalent ways to proceed in this situation. First, one could use the Dirac method introducing momenta for all variables and then going through the reduction of first and second class constraints. The other possibility is to try to diagonalise the kinetical term isolating the zero eigenvalues (or non-canonical directions) via a Darboux transformation. Here we shall follow this last possibility. 

It is a remarkable fact that the Darboux transformation that brings the kinetical term in (\ref{action/1}) to a canonical form can be performed in an explicitly Lorentz invariant way.   Indeed, consider the following
(non-linear) field redefinition of the spin connection,
\bb
w^{ab}_i = \Theta^{ab\ c}_{\ i\  j}\ p^j_c + U^{ab\ (kl) }_{\,i }
\lambda_{kl} 
\label{trans}
\ee
where $p^i_c$ and $\lambda_{kl}=\lambda_{lk}$ are 12+6=18 new independent
fields and $U$ and $\Theta$ are defined in (\ref{U},\ref{Theta}).  

Since the pair $\Theta$ and $U$ form a basis for vectors $(^{ab}_i)$, this transformation is a change of basis of the spin connection, 
\bb
w^{ab}_i  \rightarrow  (p^i_a,\lambda_{ij}),
\ee
where $p^i_a$ and $\lambda_{ij}$ are, respectively,
the components of $w^{ab}_i$ along $\Theta$ and $U$. 

The above transformation is of course invertible and one can express
$p^i_a$ and $\lambda_{ij}$ in terms of the spin connection. The formula for $p^i_a$ is simply,   
\bb
p^i_a = \Omega^{\, i\ \, j}_{a\ bc} w^{bc}_j,
\label{p}
\ee
and follows from multiplying (\ref{trans}) with $\Omega$ and using (\ref{OT}) and (\ref{OU}).  For later reference we mention here that in view of (\ref{p}) and the fact that $w^{ab}_i$ is a connection for the Lorentz group, $p^i_a$ does not transform as a vector under Lorentz rotations. The formula for $\lambda_{ij}$ is
\begin{equation}
\lambda_{ij} = \frac{-1}{8} E^m_a E^n_b \epsilon_{mn(i} w^{ab}_{j)}
\end{equation}
where $E^i_a = g^{ij} e_{j a}$ and $(ij)$ denotes symmetrization. 

Replacing the redefinition (\ref{trans}) in the action (\ref{action/1})
and using (\ref{OT}) and (\ref{OU}) we obtain,  
\bb
S[e^a_i, p^i_b, \lambda_{ij}, \omega^{a b}_{\ \ 0}, e^{a}_{0} ]
=\int dt \int_\Sigma (p^i_{c} \dot e^c_i - \omega^{a b}_{ \ \ 0} J_{a
b} - e^a_0 P_{a}), 
\label{action/3}
\ee
where the constraints have to be expressed in terms of $p^i_a$ and 
$\lambda_{ij}$ through (\ref{trans}).

The most important achievement of the above change of variables is that we
have extracted from the spin connection the part of it which is
canonically conjugate to the tetrad. Indeed, from (\ref{action/3}) we
immediately obtain the basic Poisson bracket,
\bb
\{ e^a_i(x),p^j_b(y) \}=\delta^a_b \delta^j_i \delta^{(3)}(x,y),
\label{Poisson-Bracket}
\ee
and hence, the variables $e^a_i$ and $p^i_a$ are the Darboux coordinates
of the problem. 

The components $\lambda_{ij}$, on the other hand, are not dynamical because they do not enter in the kinetical term.  This implies that the variation of (\ref{action/3}) with respect to $\lambda_{ij}$ gives six new constraints. These new constraints do not generate any new first class symmetries, instead, their only role is to determine $\lambda_{ij}$ in terms of the other canonical variables. Indeed, we shall see below that $\lambda_{ij}$ enters algebraically in the action and the equation $\partial S/\partial \lambda_{ij}=0$ can be solved in a unique way for $\lambda_{ij}$.  This should not be too surprising. After all, the $\lambda_{ij}$'s are part of the spin connection which is well known to be an auxiliary field. The relevant point here is that one can separate the part of the spin connection which is conjugate to the tetrad in an explicitly covariant form.  One can thus eliminate the $\lambda$'s from the action obtaining a new action that depends only on the tetrad and its conjugate momentum $p^i_a$. 

\subsection{Geometrical meaning of the constraints}

The action (\ref{action/3}) possesses a set constraints which come from the variation with respect to $e^a_0$ and $w^{ab}_0$ [see (\ref{const0})].
>From the tensor character of $P_a$ and $J_{ab}$ one can expect that they
generate, respectively, diffeomorphisms and local Lorentz rotations. This
is indeed the case as we now explain.

\subsubsection{Lorentz invariance}

Let us consider first the constraint $J_{ab}$ whose analysis is straightforward. It turns out that once we express the spin connection $w^{ab}_i$ in terms of the new variables by the transformation (\ref{trans}), one discovers that the constraint $J_{ab}$ does not depend on the auxiliary variables $\lambda_{ij}$.  Indeed, by direct replacement of (\ref{trans}) into (\ref{Jab}) one obtains the simple formula,
\bb
J_{a b}(e^a_i,w^{ab}_j) = - \frac{1}{2}( p_{a}^{i} e_{b i} - p_{b}^{i} e_{a i} ) +
2 \epsilon^{i j k}\epsilon_{a b c d} e^c_{j,i} e^{d}_{k}.
\label{Jab/new}
\ee
The first term in this expression is the usual (tetrad) $SO(3,1)$
generator. It turns out that the full generator (\ref{Jab/new}) also satisfy the $SO(3,1)$ algebra on the Poisson bracket (\ref{Poisson-Bracket}). The role of the second term is to remind that $p^i_a$ does not transform as a vector. The transformation induced by (\ref{Jab/new}) on $p^i_a$ is,
\bb
\delta p^i_a =  p^i_b \rho_{\ a}^b+ 
    \epsilon^{i j k} \epsilon_{abcd} e^b_j  \rho^{cd}_{,k}, 
\ee
with $\rho^{ab}=-\rho^{ba}$.  This transformation is consistent with the formula (\ref{p}) and the fact that $w^{ab}_i$ transforms as a connection. On the other hand, since the second term in (\ref{Jab/new}) is independent of $p^i_a$, the transformation induced on the tetrad is a standard $SO(3,1)$ rotation. 

Finally, we point out that since $J_{ab}$ generates a symmetry of
the action, it is of first class in the Dirac sense.   

\subsubsection{Diffeomorphism invariance}

We now analyse the constraint $P_a$.  This constraint explicitly depends on the auxiliary variable $\lambda_{ij}$. As we mention above, $\lambda_{ij}$ can be solved algebraically from its own equations of motion (see below) and therefore it does not represent any dynamical degree of freedom.  However, it is useful to keep $\lambda_{ij}$ in order to maintain the expressions short. The explicit elimination of $\lambda_{ij}$ will be studied below, before that we would like to argue here that the constraint $P_a$ does generate a symmetry of the action and therefore the time evolution is consistent. 
   
The question to be address here is whether the Poisson bracket of
$e^a_i$ and $p^i_a$ with $P_a$ are symmetries of the action. To that
end we define the smeared generator
\bb
P(\eta) = \int \eta^a P_a
\ee
and compute the Poisson brackets $\{e^a_i, P(\eta) \}$ and
$\{p^i_a,P(\eta)\}$.  

In the analysis that follows it will be important and useful to distinguish between true gauge symmetries and trivial symmetries (see
\cite{HT-book}).  Let $\phi^a$ denotes the dynamical fields in a given theory described by an action $S[\phi^a]$. It follows that the transformation $\delta \phi^a=\epsilon^{ab} (\delta S/\delta \phi^b)$ is a symmetry of the action provide $\epsilon_{ab}$ is an antisymmetric tensor, $\epsilon^{ab}=-\epsilon^{ba}$, but otherwise arbitrary. These symmetries --which by definition vanish on-shell-- are called ``trivial symmetries" because they are present in any theory that can be derived from an action. They do not give rise to constraints or conserved charges.  An interesting theorem proved in \cite{HT-book} states that any symmetry that vanishes on-shell can be expressed as a trivial symmetry. The set of true gauge symmetries is thus the quotient space of all gauge symmetries divided by the trivial symmetries. In particular, two gauge symmetries that differ by a trivial symmetry are identified in the quotient space.  

Gauge theories provide a natural example of these symmetries. Let $S[\phi^a,\rho_i]$ be the action of a gauge theory with constraints $\chi^i(\phi) :=\delta S/\delta \rho_i \approx 0$.   Now, consider a deformation of the dynamical fields $\phi^a$ given by an arbitrary combination of the constraints, $\delta \phi^a =\mu^a_i \chi^i\approx 0$ with $\mu^a_i$ an arbitrary smooth function of $\phi$. It follows that this deformation is a symmetry of the action provide the Lagrange multipliers transform as $\delta \rho_i = -\mu^a_i \delta S/\delta \phi^a\approx 0$. Note that both $\delta \phi^a$ and $\delta \rho_i$ are zero on-shell and that $\delta \phi^a$ vanishes by virtue of the equations of motion associated to $\rho_i$, while $\delta\rho_i$ vanishes by virtue of the equations associated to $\phi^a$. The minus sign in $\delta \rho_i$ accounts for the antisymmetric tensor needed in a trivial symmetry. The following  lemma will of great help for us:  {\it Any deformation of the dynamical variables which is proportional to the constraints represents a trivial symmetry of the action.}

By the direct application of the Poisson bracket   
(\ref{Poisson-Bracket}) one finds the deformation induced by $P(\eta)$ on the tetrad,
\bqn
\delta_\eta e^a_i &=& \{e^a_i,P(\eta)\}  \nonumber\\
                  &=& D_i \eta^a + 
     \Theta^{abc}_{ij} \epsilon^{jkl} \epsilon_{bcde} \eta^d T^e_{kl}
\eqn
The term $D_i \eta^a$ is, up to a term proportional to the torsion that vanishes on-shell, a diffeomorphism with a parameter $\xi^\mu = e^\mu_a \eta^a$. The second term involves also the torsion tensor and therefore it is also a trivial transformation.  In summary, the action of $P_a$ on the tetrad induces, as expected, a diffeomorphism plus a trivial transformation.

The transformation induced by $P(\eta)$ on $p^i_a$ is given by,
\bb
\delta p_{a}^{i} = - 2 \epsilon_{a b c d} \epsilon^{i m n}
\omega^{b c}_{m} D_{n}(\eta^{d}) 
+ \epsilon_{a b c d} R^{b c}_{m n} \epsilon^{i m n} \eta^{d}
- \epsilon_{b c d e} \frac{\delta \omega^{b c}_{k}(e,
\pi)}{\delta e^{a}_{i}} \eta^{d} T^{e}_{m n} \epsilon^{k m n} .
\label{delp}
\ee
The last term in this expression is zero on-shell and therefore it is a trivial transformation and we do not need to worry about it.   

In order to show that the other two terms represent a diffeomorphism acting on $p^i_a$, we recall the transformation laws under an improved  diffeomorphism with parameter $\epsilon^\mu$ on the tetrad and spin connection, 
\begin{eqnarray}
\delta e^{a}_{\mu } &=& - D_{\mu}(\epsilon^{\nu} e^{a}_{\nu}) +
T^{a}_{\mu \nu} \epsilon^{\nu} , 
\label{dif-e}\\
\delta \omega^{a b}_{\mu} &=& R^{a b}_{\mu \nu} \epsilon^{\nu}.
\label{dif-w}
\end{eqnarray}
Computing the transformation of $p^i_a$ by replacing (\ref{dif-e}) and (\ref{dif-w}) in (\ref{p}), one obtains (\ref{delp}), as expected, plus a trivial trasformation that involves the constraint $P_a=0$. 

\subsection{Elimination of the auxiliary fields}

In this section we prove that the fields $\lambda_{ij}$ are auxiliary in the sense that they can be determined algebraically from their own equations of motion.  This is easily done by projecting the constraint
$P_a$ along the normal $n^a$ and tangent vectors $e^a_i$ to the spatial
surface,
\bb
H_{\perp} = n^a P_{a}, \spc H_{i} = e^{a}_{i} P_{a},
\ee
with $n^{a}$ given by (\ref{normal}).
One can also define new Lagrange multipliers $N$ and $N^i$ by $e^a_0 = N n^a + N^i e^a_i$ and thus
\begin{equation}
e^a_0 P_a = N H_\perp + N^i H_i. 
\end{equation}

Since $\lambda_{ij}$ appears only in $P_a$, or equivalently in $H_\perp$
and $H_i$, the equations of motion associated to $\lambda_{ij}$ are
obtained by varying the combination $N H_\perp +  N^k H_k$. One obtains,
\bb
2 \sqrt{g} N \Phi^{ij} + \epsilon^{kl (i} N^{j)} e^a_k e^b_l  J_{a b} = 0,
\label{w}
\ee
where
\begin{eqnarray}
\Phi^{ij} &:=& \frac{\partial H_\perp}{\partial \lambda_{ij}} \nonumber\\
          &=& - 2 g G^{ij kl} \lambda_{kl} + 
             2 \epsilon^{kl (i} E^{j)}_{a} e^{a}_{k,l}.
\label{phi}
\end{eqnarray}
Here the parenthesis indicate symmetrization. in $(m,n)$ and
$G^{ijkl}$ is the de Witt super-metric constructed with the inverse of
the induced metric $g_{ij} = e^a_i e^b_j\eta_{ab}$, $g = det(g_{ij})$ and
$E_{a}^{i} = g^{ij}
e_{a j}$.   Since $N$ and $N^i$ are
independent Lagrange multipliers each term in (\ref{w}) has to vanish
independently.  The second term is proportional to the constraint $J_{ab}$
and thus it does not give any new equation. The first term implies $\Phi^{ij}=0$ from where we can solve $\lambda_{ij}$ as,
\bb
 \lambda_{p q} = \frac{1}{2g} G_{pq mn} \epsilon^{i j n} E^{m}_{a}
\frac{\partial e^{a}_{i}}{\partial x^{j}},  
\label{s1}
\ee
where $G^{mn pq} G_{pq rs} = \delta^{mn}_{rs}$. This finishes the proof
that the variables $\lambda_{ij}$ can be solved from its own equations of
motion. The next step is to replace (\ref{s1}) in the constraints and
write the action in terms of the canonical variables only.  Unfortunately, the expressions are complicated and not very illuminative. 

A quicker way to eliminate $\lambda_{ij}$ from $H_\perp$ and $H_i$ is implemented by the following observation.  By definition of $\Phi^{ij}$ [see (\ref{phi})], one can see that the combination,
\bb
\bar{H}_{\perp} = H_{\perp} - \frac{1}{8 \sqrt{g}} G_{mn pq}
\Phi^{mn} \Phi^{pq},
\ee
does not depend on $\lambda_{ij}$. In the same way, the combination 
\bb
\bar{H}_i = H_i  - \frac{1}{4 g} \epsilon^{r s j} e^a_r e^b_s J_{a b}\ 
  G_{ij kl}  \Phi^{kl},
\ee
does not depend on $\lambda_{ij}$ either.  They are indeed equal to $H_\perp$ and $H_i$ after $\lambda_{ij}$ is eliminated from its equations of motion. 

The last step is to compute the constraint algebra satisfied by $\bar H$,
$\bar H_i$ and $J_{ab}$.  Unfortunately, the calculations are complicated due to the intricate dependence of the constraints $\bar H_\perp$ and $\bar H_i$ on the canonical variables. One can start this calculation by computing the Poisson bracket of $P_a$ with itself. This bracket has the form (omiting indices),
\begin{equation}
\{P,P\} \sim T + T^2 + T P 
\end{equation}
where $T^a_{ij}$ are the spatial compenents of the torsion and $J$ is the generator of local Lorentz rotations. Note that $T^a_{ij}$ vanishes as a consequence of the equations of motion and since it does not have any time derivatives it is a constraint. Indeed, $T^a_{ij}=0$ are 12 equations equivalent to $\Phi^{ij}=0$ plus $J_{ab} =0$.   
The appearence of the quadratic terms $T P$ and $T^2$ suggests that the algebra of constraints is not isomorphic to the Dirac algebra but rather to its modification, first reported by Henneaux and Henneaux, Charap and Nelson (HCN) \cite{Henneaux}, which also has quadratic terms in the constraints.  One can actually expect the algebra to be isomorphic to the HCN algebra in view of a theorem proved in \cite{Clayton} which states that the integrability conditions for a consistent evolution in any tetrad theory leads to the HCN algebra.         
  
\section{Conclusions}
  
We have considered in this paper the first order form of tetrad gravity.  We have introduced a non-linear field redefinition allowing the separation of canonical and non-canonical coordinates and have analyse the structure and geometrical meaning of the constraints written in these new variables. 
This opens a new line of attack for the problem of canonical quantum gravity. Three open problems that we expect to analyse in the future are, to find a closed and explicit form for $P_a$ in terms of the tetrad and its conjugate, to prove explicitely  that the algebra of constraints is isomorphic to the HCN algebra, and to study the issue of boundary conditions and boundary terms in the new coordinates.  

\section{Acknowledgements}

During this work we have benefited from many conversations with Andy Gomberoff, Marc Henneaux, Claudio Teitelboim and, specially, Jorge Zanelli who participated in an earlier stage of this project. We also thank Steve Carlip for useful comments. This work was partially supported by the grant \# 1970150 and \# 2960007 from FONDECYT (Chile), and institutional support by a group of Chilean companies (EMPRESAS CMPC, CGE, COPEC, CODELCO, MINERA LA ESCONDIDA, NOVAGAS, ENERSIS, BUSINESS DESIGN ASS. and XEROX Chile).

\end{document}